\documentclass[aps,prl,twocolumn,amsmath,amssymb,superscriptaddress,nobibnotes]{revtex4-2}

\usepackage{amsmath,amssymb,amsfonts,amsthm}
\usepackage{bm}
\usepackage{graphicx}
\usepackage{xcolor}
\usepackage{booktabs}
\usepackage{morefloats}
\usepackage[
bookmarksnumbered,
pdfpagelabels=true,
plainpages=false,
colorlinks=true,
linkcolor=blue,
citecolor=blue,
urlcolor=blue
]{hyperref}

\begin{document}
	\title{Scalar-Wave Dispersion in Vectorial Photonic Crystals via Site-Adapted \texorpdfstring{$p$}{p} Orbitals}
	
	\author{Yan-Long Chen}
	\affiliation{School of Physics and Electronics, Hunan University, Changsha, China}
	
	\author{Kin Hung Fung}
	\affiliation{Department of Physics and State Key Laboratory of Quantum Optical Materials, The Hong Kong University of Science and Technology, Clear Water Bay, Kowloon, Hong Kong, China}
	
	\author{C. T. Chan}
	\email{phchan@ust.hk}
	\affiliation{Department of Physics and State Key Laboratory of Quantum Optical Materials, The Hong Kong University of Science and Technology, Clear Water Bay, Kowloon, Hong Kong, China}
	
	\author{Qinghua Guo}
	\email{guoqh@hnu.edu.cn}
	\affiliation{School of Physics and Electronics, Hunan University, Changsha, China}
	
	\date{\today}
	
	\begin{abstract}
		
		Electromagnetic waves are intrinsically vectorial and require description via polarization, unlike scalar fields such as acoustic pressure or electronic wavefunctions. In three dimensions, the transversality constraint further prevents any globally smooth transverse-polarization frame at the $\Gamma$ point, which would apparently rule out a simple scalar band structure for three-dimensional (3D) photonic crystals. We show here that site-adapted $p$-orbitals can realize scalar-wave dispersion: the induced band representation is isomorphic to the scalar elementary band representation up to a one-dimensional character twist, so the symmetry-enforced degeneracies and compatibility relations are the same. We demonstrate this mechanism experimentally in 3D photonic meta-crystals, where the local $p$-orbital axes adapt from site to site according to symmetry. In contrast to a fixed-polarization reduction (e.g., in 2D), our construction preserves site-polarization textures while simultaneously supporting a scalar network with one amplitude per site. Thus, it offers a pathway from vectorial photonic degrees of freedom to scalar band engineering, keeping polarization as an active design knob.
		
	\end{abstract}
	
	\maketitle
	
	\section*{Introduction}
	
	Tight-binding theory is the standard language for topological band engineering \cite{Haldane1988,KaneMele2005,Bradlyn2017TQC,Cano2018EBR,CanoBradlyn2021Review}. In its simplest form, each lattice site carries one scalar orbital (see Fig.~\ref{fig:orbital-scalarization}\textbf{a} left), and the crystal geometry enters through structure factors. This picture underlies Dirac cones \cite{Novoselov2005Graphene,Peleg2007}, flat bands \cite{Tang2011FlatBand,Mukherjee2015Lieb}, and many other fundamental topological phases \cite{Thouless1982TKNN,HaldaneRaghu2008,Wang2009,Lu2014TopologicalPhotonics,Ozawa2019TopologicalPhotonics}. A three-dimensional photonic crystal, however, is not generically described by a scalar wavefunction. In the source-free Maxwell eigenproblem, the electric field is a vector field constrained by $\nabla\cdot\bm D(\bm r)=0$, and a globally smooth transverse-polarization frame is obstructed near $\Gamma$ \cite{MoralesPerez2025,Christensen2022,WatanabeLu2018}. Consequently, a naive scalar one-orbital reduction based on a globally fixed transverse polarization is generally not a faithful reduction of Maxwell's equations \cite{MoralesPerez2025,Christensen2022}. Most practical photonic designs avoid this difficulty by using a fixed polarization in a two-dimensional geometry \cite{Khanikaev2013,Hafezi2013,WuHu2015,Noh2018ValleyHall,Shalaev2019,Ghorashi2024}, or an array of highly anisotropic waveguides in quasi-3D \cite{Rechtsman2013Floquet,Plotnik2014,Sun2024FloquetSkin}. Such strategies are powerful, but they do not provide a general route from a fully vectorial 3D Maxwell eigenproblem to a desired scalar tight-binding Hamiltonian. Recent regularization and transversality-enforced tight-binding approaches address the $\Gamma$-point obstruction using auxiliary apolar or longitudinal modes \cite{MoralesPerez2025,Christensen2022}. Here, we instead isolate a one-dimensional sector of the cavity dipolar manifold---while preserving measurable adaptive local polarization---and find that this sector obeys an $s$-orbital tight-binding dispersion, characteristic of scalar behavior (see Fig.~\ref{fig:orbital-scalarization}\textbf{a} right).
	
	A localized dipolar resonance of a photonic cavity is the electromagnetic analogue of a $p$-orbital triplet (see Fig.~\ref{fig:orbital-scalarization}\textbf{b}) \cite{Zhang2023PhotonicPOrbital,Kozon2024,Milicevic2017}. At a Wyckoff position with a site-symmetry group, this three-dimensional polar-vector representation can split into a one-dimensional sector and an orthogonal two-dimensional sector. Site symmetry forbids onsite mixing between inequivalent local irreducible representations. If an irreducible representation occurs more than once, site symmetry alone does not select a unique copy but the geometry and channel design can be used to further isolate the desired linear combination. A symmetry-compatible channel network can thus make the projected hopping block diagonal, leaving the selected one-dimensional sector closed throughout the Brillouin zone (see Fig.~\ref{fig:orbital-scalarization}\textbf{c}). This sector forms an elementary band representation (EBR) induced from the corresponding site-symmetry irrep \cite{Bradlyn2017TQC,Cano2018EBR,CanoBradlyn2021Review}. It is described by one scalar coefficient per site even though the underlying eigenfield remains a vector field with a site-dependent polarization direction (see Figs.~\ref{fig:orbital-scalarization}\textbf{d} and~\ref{fig:orbital-scalarization}\textbf{e}).
	
	\section*{Results}
	
	\subsection*{Orbital-selective scalarization mechanism}
	
	Let $i$ and $j$ be adjacent resonators connected by a channel with unit bond vector $\bm d_{ij}$, and let $|p_i\rangle$ and $|p_j\rangle$ denote the normalized selected local $p$-orbital states. In the full local $p$-orbital space, the bond hopping operator takes the standard Slater--Koster form \cite{SlaterKoster1954},
	\begin{equation}
		\hat T_{ij}=
		V_{pp\sigma}|\bm d_{ij}\rangle\langle\bm d_{ij}|
		+V_{pp\pi}\left(I_3-|\bm d_{ij}\rangle\langle\bm d_{ij}|\right),
		\label{eq:p-hopping-main}
	\end{equation}
	and its projection onto the selected one-dimensional local sector is
	\begin{equation}
		t_{ij}=\langle p_i|\hat T_{ij}|p_j\rangle.
		\label{eq:projected-hopping-main}
	\end{equation}
	The effective nearest-neighbor Hamiltonian is
	\begin{equation}
		\hat H=\sum_{\langle i,j\rangle}
		\left(t_{ij}c_i^\dagger c_j+\mathrm{h.c.}\right),
		\label{eq:one-p-orbital}
	\end{equation}
	where $c_i^\dagger$ creates an excitation in the selected localized mode at site $i$. This Hamiltonian coincides with the $s$-orbital one when $t_{ij}$ is isotropic. Thus, scalarization amounts to requiring equal $t_{ij}$ on symmetry-equivalent bonds, closely resembling the isotropy of the $s$ orbital (see Supplemental Material Sec.~S1). Although this condition may seem overly restrictive, we surprisingly find it to be quite general. Mackey's tensor product theorem offers the key insight, providing a rigorous group-theoretic connection between local orbital symmetries and global band structures \cite{Mackey1952,Fakler1973}.
	
	To make this precise, let $\mathcal{G}$ be a space group with translation subgroup $\mathcal{T}\cong\mathbb{Z}^3$ and point group $P = \mathcal{G}/\mathcal{T}$. For a given Wyckoff position with site-symmetry group $H$ (the stabilizer of a representative site), a band representation $\rho \uparrow\mathcal{G}$ is obtained by inducing a local representation $\rho$ of $H$ to the full space group \cite{Zak1982,Bradlyn2017TQC,Cano2018EBR}. If the local $p$-orbital representation $\rho_p$ is the restriction of a one-dimensional representation $\chi$ of the point group $P$, then Mackey's theorem yields the isomorphism \cite{Mackey1952,Fakler1973,Zak1982}
	\begin{equation}
		\mathcal{B}_p \cong \mathcal{B}_s \otimes \tilde{\chi},
		\label{eq:band-representation-isomorphism}
	\end{equation}
	where $\mathcal{B}_s$ is the band representation induced from the trivial ($s$-orbital) representation and $\tilde{\chi}$ is the lift of $\chi$ to $\mathcal{G}$ defined by $\tilde{\chi}(\{R\mid\bm t\}) = \chi(R)$. Because this isomorphism holds for the entire space group, it restricts to every little group $\mathcal{G}_{\bm k}$, giving $\mathcal{B}_p^{\bm k} \cong \mathcal{B}_s^{\bm k} \otimes \tilde{\chi}|_{\mathcal{G}_{\bm k}}$ for all $\bm k$ in the Brillouin zone. Thus the two band structures share identical degeneracies and splitting patterns throughout the entire Brillouin zone, differing only in their symmetry labels by the factor $\tilde{\chi}$ (see Supplemental Material Secs.~S2--S4).
	
	The local $p$-orbital space decomposes as $\rho_p \oplus \rho_\perp$, where $\rho_p$ denotes a one-dimensional irrep and $\rho_\perp$ is the complementary sector. When $\rho_p=\mathbf{1}$ is trivial, the resulting band is automatically identical to that of the $s$-orbital. If $\rho_p$ is nontrivial but satisfies the character-extension condition, it has
	the same scalar band connectivity and symmetry-enforced degeneracies and scalar dispersion follows when the projected hopping block is matched to the scalar model. We now present three examples that illustrate the different scenarios.
	
	\begin{figure}[t]
		\includegraphics[width=1.0\columnwidth]{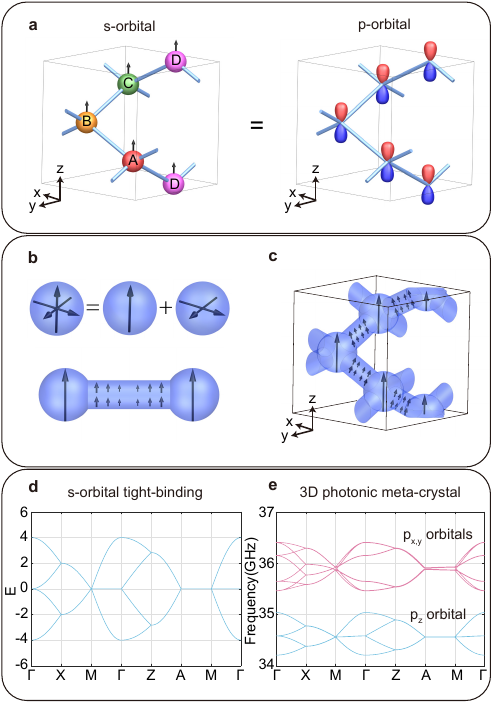}
		\caption{Orbital-selective scalarization of vectorial photonic modes. \textbf{a} Left: tight-binding model on the Wyckoff positions (4a) of space group No.~92. Right: corresponding $p$-orbital construction obtained by replacing the $s$-orbital at each Wyckoff position with a selected local $p$-orbital. Within our framework, the $p$-orbital Hamiltonian reproduces the scalar $s$-orbital tight-binding model. \textbf{b} A spherical cavity supports a dipolar triplet which generates a three-component dipolar p-like manifold and symmetry-compatible anisotropy or coupling to the channel network can be used to isolate one component from the other two. \textbf{c} In a 3D network, the 4a Wyckoff positions [$\mathrm{A}=(-\tfrac{1}{4}, \tfrac{1}{4}, -\tfrac{1}{4})$, $\mathrm{B}=(\tfrac{1}{4}, \tfrac{1}{4}, 0)$, $\mathrm{C}=(\tfrac{1}{4}, -\tfrac{1}{4}, \tfrac{1}{4})$, $\mathrm{D}=(-\tfrac{1}{4}, -\tfrac{1}{4}, \tfrac{1}{2})$] of SG~92 define symmetry-related local polarization axes. \textbf{d} Target scalar $s$-orbital tight-binding dispersion. \textbf{e} Full-wave spectrum of the SG~92 photonic meta-crystal. The selected $p_z$-derived four-band manifold (cyan) follows the scalar model, while the complementary $p_{x,y}$-derived eight-band sector (pink) remains separated. The channel cutoff is $f_c=39.96\,\mathrm{GHz}$.}
		\label{fig:orbital-scalarization}
	\end{figure}
	
	\begin{figure}[t]
		\includegraphics[width=1.0\columnwidth]{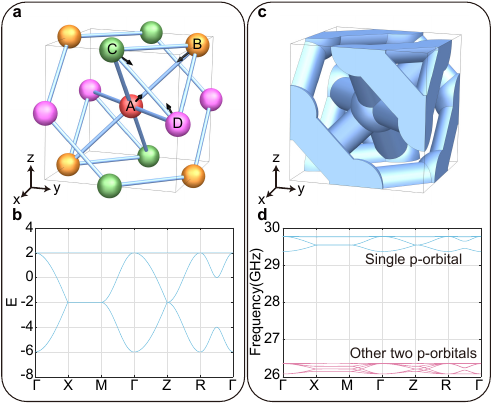}
		\caption{Cubic orbital scalarization manifold in SG~224. \textbf{a} Four-site scalar tight-binding network on the 4b Wyckoff orbit [$\mathrm{A}=(0,0,0)$, $\mathrm{B}=(0,\tfrac{1}{2},\tfrac{1}{2})$, $\mathrm{C}=(\tfrac{1}{2},0,\tfrac{1}{2})$, and $\mathrm{D}=(\tfrac{1}{2},\tfrac{1}{2},0)$]. \textbf{b} Scalar $s$-orbital dispersion with three-dimensional flat bands. \textbf{c} Photonic meta-crystal, with blue regions denoting air channels and the remaining volume modeled as PEC. \textbf{d} Full-wave spectrum. The selected four-band $A_{2u}\uparrow\mathcal{G}$ EBR is highlighted in cyan and is separated from the transverse $E_u$-derived sector. The channel cutoff is $f_c=35.16\,\mathrm{GHz}$.}
		\label{fig:sg224}
	\end{figure}
	
	\subsection*{Nonsymmorphic tetragonal example: \texorpdfstring{{\boldmath$\mathrm{P}4_{1}2_{1}2$}}{P4(1)2(1)2} (No.~92)}
	
	The first implementation employs the nonsymmorphic tetragonal space group {\boldmath$\mathrm{P}4_{1}2_{1}2$} (No.~92). The four resonators occupy the $4a$ Wyckoff orbit (see Fig.~\ref{fig:orbital-scalarization}\textbf{a} left), which has $..2$ site symmetry. The site-symmetry group $H=C_2$ has its twofold axis along $[110]$ \cite{Aroyo2006BCSI,Aroyo2006BCSII}. The selected $p_z$ orbital, being transverse to this axis, is odd under the twofold rotation and thus transforms as the nontrivial $C_2$ irrep $B$. This local irrep is the restriction of the irrep $A_2$ of point group $P=D_4$, so the condition of Mackey's theorem is satisfied. In the EBR description, the four cyan bands correspond to $B \uparrow\mathcal{G}$ and are gapped from the remaining $p$-derived bands (see Fig.~\ref{fig:orbital-scalarization}\textbf{e}).
	
	At $\Gamma$, the point group is $D_4$ \cite{Aroyo2006BCSII}. The $s$-orbital-induced band representation reduces to
	\begin{equation}
		\rho_s^\Gamma = A_1 \oplus B_2 \oplus E,
		\label{eq:sg92-s-gamma}
	\end{equation}
	while its $p$-orbital counterpart reduces to
	\begin{equation}
		\rho_p^\Gamma = A_2 \oplus B_1 \oplus E.
		\label{eq:sg92-p-gamma}
	\end{equation}
	These two decompositions are not identical but are related by tensor multiplication with the $D_4$ irrep $A_2$, as $A_2 \otimes \rho_s^\Gamma \cong \rho_p^\Gamma$. Hence, at $\Gamma$, flipping the local $C_2$ character from $+1$ to $-1$ is equivalent to an overall multiplication by $A_2$, exactly as predicted by the general theorem. As seen in Figs.~\ref{fig:orbital-scalarization}\textbf{d} and~\ref{fig:orbital-scalarization}\textbf{e}, this confirms that the two band representations possess identical degeneracies and splitting patterns, differing only in symmetry labels---a direct manifestation of Mackey's theorem discussed above. While the above demonstration is given at $\Gamma$ for simplicity, the isomorphism is global---it remains valid for all $\bm k$ throughout the first Brillouin zone, as guaranteed by the restriction to every little group $\mathcal{G}_{\bm k}$.
	
	\subsection*{Cubic flat-band example: \texorpdfstring{{\boldmath$\mathrm{Pn}\bar{3}\mathrm{m}$}}{Pn-3m} (No.~224)}
	
	The second example is a centrosymmetric cubic meta-crystal in space group $\mathrm{Pn}\bar{3}\mathrm{m}$ (No.~224). The four resonators occupy the 4b Wyckoff positions (see Fig.~\ref{fig:sg224}\textbf{a}). These sites have local $.\bar{3}\mathrm{m}$ site symmetry, with local axes along $[111]$ (see Supplemental Fig.~S1) \cite{Aroyo2006BCSI,Aroyo2006BCSII}. The local polar-vector representation decomposes as $A_{2u}\oplus E_u$ of $H=D_{3d}$: the component along the local threefold axis transforms as $A_{2u}$, while the two transverse components transform as $E_u$. The selected $A_{2u}\uparrow\mathcal{G}$ EBR is a four-band manifold. The local $A_{2u}$ irrep extends to the one-dimensional $A_{2u}$ of $O_h$, so this EBR is globally related to the $A_{1g}$-induced scalar EBR by a character twist. Because the local axes differ among $\mathrm{A}$, $\mathrm{B}$, $\mathrm{C}$, and $\mathrm{D}$, this manifold cannot be described by a fixed global polarization (see Supplemental Figs.~S1 and S2). Nonetheless, the system continues to obey the scalar four-site tight-binding model (Fig.~\ref{fig:sg224}\textbf{b}). As seen in Figs.~\ref{fig:sg224}\textbf{c} and~\ref{fig:sg224}\textbf{d}, below $f_c = 35.16\,\mathrm{GHz}$, the cyan bands form an isolated scalar-like manifold, while the transverse $p$-orbital sector constitutes a separate eight-band manifold. This demonstrates that scalarization is possible without requiring every site to share the same global polarization axis.
	
	A similar analysis applies at $\Gamma$ for the cubic space group, where the point group is $O_h$ \cite{Aroyo2006BCSII}. For an $s$-orbital (trivial, even under inversion), the induced band representation reduces to
	\begin{equation}
		\rho_{A_{1g}}^\Gamma = A_{1g} \oplus T_{2g},
		\label{eq:sg224-s-gamma}
	\end{equation}
	while for a $p$-orbital (odd under inversion, transforming as $A_{2u}$), it reduces to
	\begin{equation}
		\rho_{A_{2u}}^\Gamma = A_{2u} \oplus T_{1u}.
		\label{eq:sg224-p-gamma}
	\end{equation}
	These two decompositions are related by $A_{2u} \otimes \rho_{A_{1g}}^\Gamma \cong \rho_{A_{2u}}^\Gamma$, which follows from the multiplication rules $A_{2u} \otimes A_{1g} \cong A_{2u}$ and $A_{2u} \otimes T_{2g} \cong T_{1u}$. Thus, at $\Gamma$, the $A_{2u}$ character---flipping parity and interchanging $T_2$ and $T_1$---acts as an overall multiplication by the $O_h$ irrep $A_{2u}$. By Mackey's theorem, this equivalence holds for all $\bm k$ in the Brillouin zone, ensuring identical degeneracies and splitting patterns between the two band structures, with only symmetry labels differing.
	
	\subsection*{Chiral example and experiment: \texorpdfstring{{\boldmath$\mathrm{P}2_{1}3$}}{P2(1)3} (No.~198)}
	
	\begin{figure}[t]
		\includegraphics[width=1.0\columnwidth]{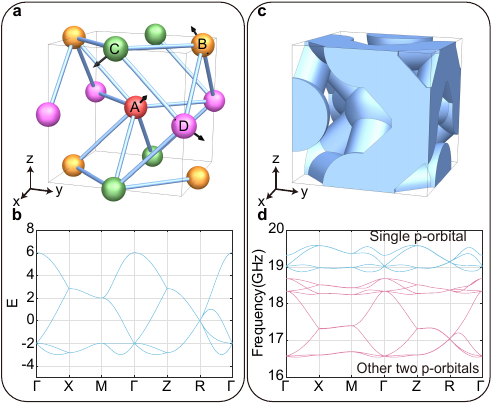}
		\caption{Chiral orbital scalarization manifold in SG~198. \textbf{a} Four-site scalar tight-binding network on the 4a Wyckoff orbit [$\mathrm{A}=(0,0,0)$, $\mathrm{B}=(0,\tfrac{1}{2},\tfrac{1}{2})$, $\mathrm{C}=(\tfrac{1}{2},0,\tfrac{1}{2})$, and $\mathrm{D}=(\tfrac{1}{2},\tfrac{1}{2},0)$]. \textbf{b} Scalar band structure with symmetry-enforced topological degeneracies. \textbf{c} Photonic meta-crystal. \textbf{d} Full-wave spectrum. The selected four-band $A\uparrow\mathcal{G}$ EBR is highlighted in cyan; the transverse $E$-derived sector is shown in pink. The channel cutoff is $f_c = 24.41\,\mathrm{GHz}$.}
		\label{fig:sg198}
	\end{figure}
	The third example uses the chiral space group $\mathrm{P}2_{1}3$ (No.~198). The four resonators occupy the $4a$ Wyckoff orbit (see Fig.~\ref{fig:sg198}\textbf{a}), whose site-symmetry group is $\mathrm{.3.}\cong C_3$ \cite{Aroyo2006BCSI,Aroyo2006BCSII}, with the local axes oriented along $[111]$ (see Supplemental Figs.~S1 and S2). Over the real numbers, the local polar-vector representation decomposes as $A\oplus E$. The longitudinal component along the local threefold axis transforms as the trivial irrep $A$, whereas the two transverse components form the two-dimensional real representation $E$. Because the local threefold axes are site dependent, the selected $A\uparrow\mathcal{G}$ manifold is not a fixed global-polarization sector. Since $A$ is the trivial local irrep of $C_3$, the induced band representation $A \uparrow\mathcal{G}$ is explicitly isomorphic to the band representation of the corresponding scalar $s$ orbital (see Fig.~\ref{fig:sg198}\textbf{b}). As shown in Figs.~\ref{fig:sg198}\textbf{c} and~\ref{fig:sg198}\textbf{d}, they share the same symmetry-enforced topological triple and quadruple degeneracies \cite{Bradlyn2016,Tang2017,Zhang2018DoubleWeyl}. This corresponds to the trivial case, where Mackey's theorem is not even required, as both the $p$-orbital and $s$-orbital share the same irreducible representations already at the site-symmetry-group level. However, Mackey's theorem becomes truly indispensable for understanding the transverse sector---the remaining eight-band manifold. At $\Gamma$ with point group $T$, the $s$-orbital-induced band representation (trivial local $A$ irrep of $C_3$) reduces to
	\begin{equation}
		\rho_s^\Gamma = A \oplus T,
		\label{eq:sg198-s-gamma}
	\end{equation}
	whereas the $p_\perp$-orbital-induced band representation reduces to
	\begin{equation}
		\rho_{p_\perp}^\Gamma = E \oplus 2T.
		\label{eq:sg198-pperp-gamma}
	\end{equation}
	These two decompositions are related through the tensor-product relations $E\otimes A\cong E$ and $E\otimes T\cong 2T$, from which we obtain
	\begin{equation}
		E \otimes \rho_s^\Gamma
		\cong \rho_{p_\perp}^\Gamma.
		\label{eq:sg198-tensor-relation}
	\end{equation}
	Thus, Mackey's theorem provides the essential framework for understanding the transverse sector, showing that even a two-dimensional local $C_3$ representation admits a tensor-product structure connecting it to the scalar $s$-orbital band.
	
	Despite the triviality of the $p$-orbital irrep, the chiral SG~198 structure generates chiral polarization textures that are absent from scalar wavefunctions but are intrinsic to vectorial photonic meta-crystals. They thus point toward a possible route for site-adaptive polarization generation and detection without a fixed-axis polarizer.
	
	We fabricated the SG~198 meta-crystal by 3D printing. The sample contains $15 \times 15 \times 7$ unit cells, with an overall size of approximately $300 \times 300 \times 140\,\mathrm{mm}^3$ and a lattice constant $a=20\,\mathrm{mm}$. A fixed electric dipole antenna excites the sample from the side, while a second dipole antenna scans the top surface to measure the amplitude and phase of the electromagnetic field (see Supplemental Fig.~S3). Fourier transformation of the real-space scan reconstructs the surface-projected momentum spectrum over the surface Brillouin zone, with $k_x,k_y\in\left[-\tfrac{\pi}{a},\tfrac{\pi}{a}\right]$.
	
\begin{figure}[htbp!]
		\includegraphics[width=1.0\columnwidth]{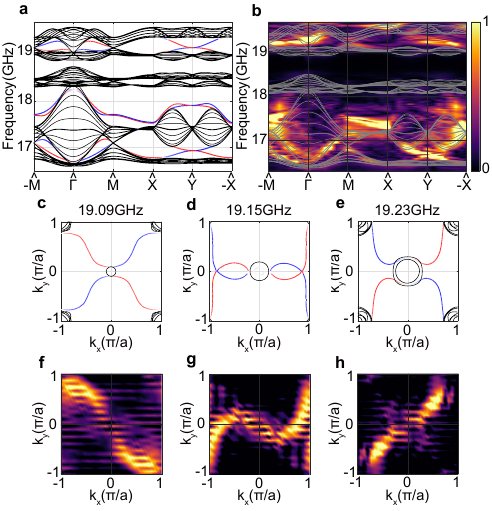}
		\caption{Experimental observation of the orbital-selective scalar-like manifold in the SG~198 meta-crystal. \textbf{a,b} Numerically simulated and experimentally reconstructed bulk- and surface-projected spectra on the (001) termination. Red and blue indicate modes localized on opposite surfaces in the calculation. \textbf{c--e} Simulated iso-frequency contours at $19.09\,\mathrm{GHz}$, $19.15\,\mathrm{GHz}$, and $19.23\,\mathrm{GHz}$. \textbf{f--h} Corresponding experimentally measured contours.}
		\label{fig:sg198-experiment}
	\end{figure}

	The measured surface spectrum agrees with the slab calculation. Figure~\ref{fig:sg198-experiment}\textbf{a} shows the calculated projected spectrum on the (001) termination, and Fig.~\ref{fig:sg198-experiment}\textbf{b} shows the experimentally reconstructed spectrum. The frequency positions and shapes of the surface-state branches are reproduced well; the small offsets are consistent with the expected fabrication tolerances. Figures~\ref{fig:sg198-experiment}\textbf{c--e} and~\ref{fig:sg198-experiment}\textbf{f--h} compare simulated and measured iso-frequency contours at $19.09\,\mathrm{GHz}$, $19.15\,\mathrm{GHz}$, and $19.23\,\mathrm{GHz}$. The observed contours confirm the predicted surface-state signatures of the chiral orbital-selective manifold. We see that the meta-crystal perfectly reproduces these surface-state branches of the target tight-binding model. Because the electromagnetic fields are vectorial, these surface states also encode site- and momentum-dependent polarization textures. Near degeneracy points, the scalar amplitudes combine different local axes and can form chiral electromagnetic states, providing a polarization function that is unavailable in a purely scalar wave model.
	
	\section*{Discussion}
	
	The selected EBR and the complementary $p$-derived sector remain energetically separated throughout the Brillouin zone. This condition naturally realizes a classical analog of electronic half-metals \cite{deGroot1983,Katsnelson2008}. In an electronic half-metal, one spin channel (say, spin-up) exhibits metallic conductivity, while the other (spin-down) remains insulating or semiconducting, leading to complete spin polarization at the Fermi level \cite{Park1998HalfMetal}. In our photonic system, the role of spin is played by the polarization degree of freedom: the selected $p$-orbital manifold---transforming according to a specific local irrep---supports dispersive modes (the ``metallic'' channel), whereas the complementary $p$-derived sector is fully gapped (the ``insulating'' channel). The site-symmetry prohibition of onsite mixing guarantees that the two polarization channels remain orthogonal and cannot hybridize at any symmetry point, while the global gap ensures that this separation persists across the entire Brillouin zone, preventing accidental band crossings. Crucially, unlike electronic half-metals, which rely on quantum correlations or spin-orbit interactions, our implementation is purely symmetry-driven and does not require fine-tuning of material parameters (see Supplemental Material Sec.~S5 and Fig.~S4). This demonstrates that half-metal-like behavior is not exclusive to quantum systems but can be engineered in classical vectorial photonic platforms.
	
	The method can be generalized systematically to other space groups. For each group, one searches Wyckoff positions whose polar-vector site representation contains a one-dimensional irrep. The key condition is that this irrep must be extendable to a one-dimensional representation of the full space group; whenever this holds, Mackey's theorem ensures the desired isomorphism to the scalar $s$-orbital band representation. The induced EBR from that irrep is the candidate scalar-like band sector. One then designs a symmetry-compatible bond network and tunes the geometry so that the selected EBR is spectrally isolated below the channel cutoff. The framework provides a systematic search procedure, although not every Wyckoff position and not every space group will realize an isolated scalar-like $p$ sector.
	
	This mechanism is distinct from conventional polarization filtering. A filter selects a fixed laboratory polarization and rejects the orthogonal one. Here, the selected ``polarization'' is a local orbital axis defined by the space group and the Wyckoff orbit. These local axes can rotate from site to site, and their relative phases can combine into momentum-dependent, even chiral, polarization textures \cite{Bliokh2015SpinOrbit,Yang2023Skyrmion,Zhen2014BIC,Zhang2018PolarizationVortex,Doeleman2018PolarizationVortex,Wang2025SpinOrbitLocking}. When the projected vectorial Hamiltonian is matched to a target scalar network, it reproduces the corresponding bulk dispersion and symmetry-enforced degeneracies. The advantage, therefore, lies not only in reproducing scalar band phenomena but also in controlling polarization textures in three-dimensional photonic crystals. Here, ``adaptive'' means that the selected local orbital axis follows the symmetry-related orientation of each Wyckoff site rather than a common laboratory polarization.

	This localized-orbital construction circumvents the obstruction---encountered in the reduced description near $\Gamma$---to a globally smooth transverse-polarization frame. The scalar amplitude is an EBR coefficient for localized resonator modes, not a free-space transverse-polarization basis (i.e., not at zero frequency). The full Maxwell field remains vectorial and satisfies $\nabla\cdot\bm D(\bm r)=0$, yet the effective scalar Hamiltonian correctly describes the symmetry-selected local-orbital sector.

	In conclusion, we have shown that scalar tight-binding band engineering can be embedded in fully vectorial three-dimensional photonics by using symmetry-selected $p$-orbital elementary band representations. Both trivial and nontrivial one-dimensional site-symmetry irreps can yield a one-amplitude-per-site scalar Hamiltonian, although they generally induce distinct EBRs and distinct symmetry labels. At suitable Wyckoff positions, the local polar-vector representation contains a one-dimensional channel, while site symmetry together with a space-group-compatible channel network can isolate the selected channel from the complementary $p$-derived subspace. When the selected channel is preserved by space-group-compatible hopping, its projected Hamiltonian reproduces the target scalar $s$-orbital band dispersion, while the electromagnetic eigenfields retain adaptive local polarization. The three examples and the microwave experiment demonstrate that the mechanism is not tied to a single geometry, inversion symmetry, or a fixed global polarization. This establishes a route from real-space crystallographic symmetry to scalar-like band design and site-momentum-dependent polarization control in 3D photonic meta-crystals.
	
	\section*{Acknowledgments}
	
	This work was supported by the National Natural Science Foundation of China
	(NSFC) under Grant Nos.~12374346 and 12322412. Work in Hong Kong is supported by
	the Research Grants Council of Hong Kong (16310422) and the State Key Laboratory
	of Quantum Optical Materials.
	
	\bibliography{MainText}
	
\end{document}